\documentclass[useAMS,usenatbib]{mn2e}
\input{psfig}

\def\ie{\mbox{\'{\i}}}

\def\cygxu{\mbox{Cygnus X-1}}

\def\grs{\mbox{GRS 1915+105}}

\def\gx{\mbox{GX 339-4}}

\def\ss{\mbox{SS 433}}

\def\xtejodh{\mbox{XTE J1118+480}}

\def\xtejdhcn{\mbox{XTE J1859+226}}

\def\cmmoinsdeux{\mbox{ cm}^{-2}}

\def\kpc{\mbox{ kpc}}

\def\Msol{\mbox{ }M_{\odot}}
\def\Rsol{\mbox{ }R_{\odot}}

\def\mag{\mbox{ magnitude}}

\def\keV{\mbox{ keV}}

\def\ergs{\mbox{ erg\,s}^{-1}}

\def\deg{^{\circ}}

\def\asec{^{\prime \prime}}

\def\nh{N_{\rm H}}

\def\ltsima{\; \buildrel < \over \sim \;}
\def\simlt{\lower.5ex\hbox{\ltsima}}            
\def\gtsima{\; \buildrel > \over \sim \;}
\def\simgt{\lower.5ex\hbox{\gtsima}}            

\usepackage{amssymb,natbib}
\begin{document}

%
   \title[Multiwavelength observations of $\xtejodh$]
{Multiwavelength observations revealing the evolution of
the outburst of the black hole $\xtejodh$ \\
}
%
   \author[S. Chaty et al.]
{S.~Chaty $^{1,2,3}$, C.A.~Haswell $^3$, J.~Malzac $^{4,5}$, 
	R.I.~Hynes $^{6,7}$, C.R.~Shrader $^8$, W.~Cui $^9$
        \\ 
$^1$ {Universit\'e Paris 7, F\'ed\'eration APC, 2 place Jussieu, 75005 Paris, 
France\thanks{chaty@cea.fr}} \\
$^2$ {Service d'Astrophysique,
DSM/DAPNIA/SAp, CEA-Saclay, Bat. 709, L'Orme des Merisiers
F-91 191 Gif-sur-Yvette, Cedex, France} \\
   $^3$ {Department of Physics and Astronomy, 
The Open University, Walton Hall, 
Milton Keynes, MK7 6AA, United Kingdom} \\
$^4$ {Institute of Astronomy, Madingley Road, Cambridge, CB3 0HA United Kingdom} \\
$^5$ {Osservatorio Astronomico di Brera, Via Brera, 28, 20121, Milano, 
  Italy} \\
$^6$ {Astronomy department, The University of Texas at Austin, 1 University Station C1400, Austin, Texas 78712-0259, USA} \\
$^7$ {Department of Physics and Astronomy, University of Southampton,
Southampton, SO17 1BJ, UK} \\
$^8$ {Laboratory for High-Energy Astrophysics, 
NASA Goddard Space Flight Center, Greenbelt, MD 20771, USA} \\
$^9$ {Department of Physics, Purdue University, 1396 Physics Building, 
West Lafayette, IN 47907-1396, USA} \\
}
   \date{Received date; accepted date}
   \pubyear{2001} \volume{000} \pagerange{1} \twocolumn

   \maketitle \label{firstpage}

   \begin{abstract}
We report multiwavelength observations of the soft
X-ray transient (SXT) $\xtejodh$, which we
observed with UKIRT, {\it HST}, {\it RXTE}, {\it EUVE} and many
other instruments and facilities.
Adding radio (Ryle Telescope, VLA), sub-millimeter (JCMT)
and X-ray ({\it Chandra} and {\it SAX}) data from the literature, we
assembled the most complete spectral energy distribution (SED) 
of this source yet published.
We followed the evolution of this source for 1 year, including 6 observations
performed during the outburst, and one observation at the end
of the outburst.
Because of $\xtejodh$'s unusually high galactic latitude, 
it suffers from very low extinction, and its SED is near-complete, 
including EUV (Extreme Ultraviolet) observations.
$\xtejodh$ exhibits an unusually low low-hard state
(estimated inner radius $350R_s$) 
and a strong non-thermal contribution in the
radio to optical domain, which is likely to be due to synchrotron emission.
We discuss the interstellar column density and show that
it is low, between 0.80 and $1.30 \times 10^{20} \cmmoinsdeux$.
We analyse the evolution of the SED 
during the outburst, including the contributions from the companion star, 
the accretion disc, the outflow, and
relating irradiation and variability of the source
in different bands to the SED.
We find no significant 
spectral variability during the outburst evolution, consistent with
the presence of a steady outflow.
Analysis of its outflow to accretion energy ratio suggests
that the microquasar $\xtejodh$ is analogous to radio-quiet quasars.
This, combined with the inverted spectrum from radio to optical, makes
$\xtejodh$ very similar to other microquasar sources, e.g. 
$\grs$ and $\gx$ in their low/hard state.
We model the high-energy emission with a hot disc model, and 
discuss different accretion models for
$\xtejodh$'s broad band spectrum.

   \end{abstract}

%

\section{Introduction} \label{introduction}

Soft X-ray Transients (SXTs), also called X-ray novae, 
are a class of low mass X-ray binaries (LMXBs).
Among this class of sources,
more than 70\% are thought to contain black holes \citep{charles:1998}.
The compact object accretes
matter through an accretion disc 
from a low-mass star via Roche lobe overflow. The history
of these sources is characterised by long periods of quiescence, 
typically lasting decades, punctuated by very dramatic 
outbursts. SXTs are usually discovered in X-rays, 
but outbursts are visible at every wavelength and in particular are 
often accompanied by radio activity. 
A typical outburst is characterized by 
soft X-ray emission dominated by thermal emission from the hot inner
accretion disc, and optical/UV emission produced by reprocessing of X-rays.

\begin{figure}
\centerline{\psfig{file=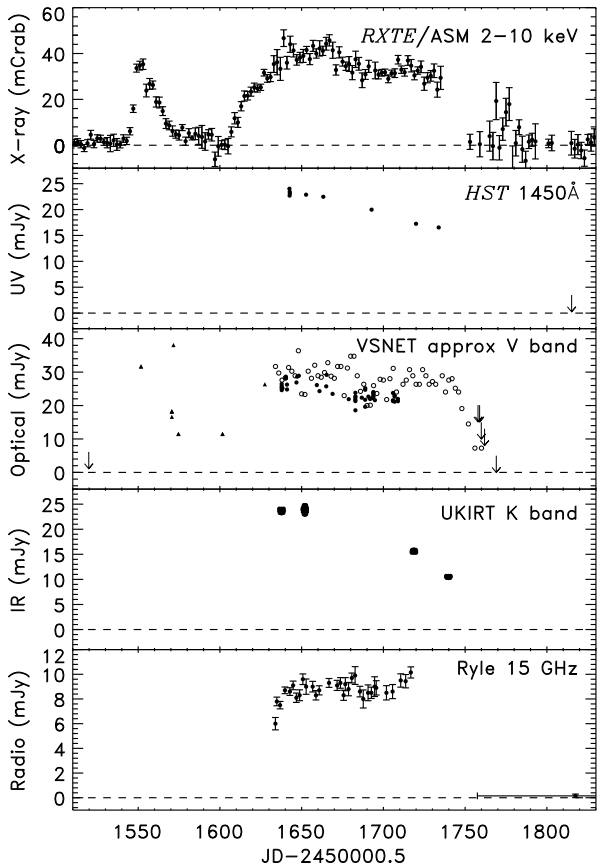,angle=0.,width=9.cm}}
\caption[]{\label{1118_lc} 
Multi-wavelength lightcurve of $\xtejodh$ in 2000:
the 6 epochs of our observations are indicated by the {\it HST} points.
Top panel shows {\it RXTE/ASM} 2--10 keV count rates in two-day averages. 
 Periods with too few
dwells for a precise measurement have been excluded for clarity.  
Second panel is
HST/STIS far-UV fluxes derived from E140M spectroscopy and an upper
limit from the late G140L observation. Middle panel shows VSNET
optical 
measurements
({\sl http://www.kusastro.kyoto-u.ac.jp/vsnet/Summary/j1118.html}). 
Open circles are two-day averages of visual estimates.  Solid
circles are individual electronic measurements.  Triangles are
pre-discovery photographic measurements.  Arrows are upper limits from
all of these sources.  The fourth panel shows K band fluxes measured
by UKIRT. The lower panel shows 15 GHz radio data from the Ryle
Telescope ({\sl http://www.mrao.cam.ac.uk/$\sim$guy/J1118+480/J1118480.list}).
The last point is based on the average flux from
2000 August to November.
}
\end{figure}

One such source, $\xtejodh$, was discovered by {\it RXTE} on 2000 March 29 
at the galactic coordinates ($l,b$) = ($157.62\deg$,$+62.32\deg$)
\citep{remillard:2000} as a weak (39 mCrab), slowly rising X-ray source, 
the post-analysis revealing
an outburst in January 2000, with a similar brightness.
The outburst history of the source is shown in Fig. \ref{1118_lc}.

The optical counterpart of $\xtejodh$ in outburst was a 13th magnitude star, 
coincident with a 18.8 mag
object in the USNO catalog \citep{uemura:2000}.
Its optical spectrum was typical
of X-ray novae in outburst \citep{garcia:2000}.
An infrared counterpart was rapidly identified thereafter \citep{chaty:2000}.
Taking the X-ray and optical fluxes given above, it is clear that
this system exhibits a very low X-ray to optical flux ratio
of 5 (in $f_{\nu}$, see \citealt{remillard:2000}; \citealt{uemura:2000}),
compared to a typical value of 500 for SXTs in outburst
(see e.g. \citealt{tanaka:1996}).
A weak 4.1 hr (0.17082 d) photometric modulation 
was rapidly discovered \citep{cook:2000}, associated
with the orbital period of the source, 
or possibly a superhump period \citep{uemura:2000}. 
$\xtejodh$ exhibits therefore 
the shortest orbital period among the black hole candidates.
\citet{wagner:2001} took spectra of this source near quiescence and
observed a period 0.5\% smaller than during the outburst. They 
suggested that this difference was due to the presence of 
superhump modulations during the outburst, later confirmed
by \cite{zurita:2002}.
Flickering with an amplitude of $\sim 0.4$ mag,
and also a quasi-periodic oscillation (QPO) 
at 10 s, was observed in the optical, in the UV
\citep{haswell:2000c} and also in X-rays \citep{revnivtsev:2000b}, 
with an evolving frequency \citep{wood:2000}.
A faint radio counterpart was detected at 6.2 mJy, but no jet feature
could be spatially resolved (see \citealt{fender:2001b}).
Optical observations of the source in quiescence led to the
determination of a large value of the mass function, 
f(M) $= 5.9 \pm 0.4 \Msol$, 
suggesting that the compact object is a black hole (\citealt{wagner:2001}; 
McClintock et al. 2001a). 

$\xtejodh$'s location at an unusually high galactic latitude ($b=+62 \deg$)
in the direction to the Lockman Hole
implies that there is a very low absorption along the line of sight
of the source. \citet{garcia:2000} estimated {\it E(B-V)} $\sim 0.024$, 
leading to a column density of $N_{H} \sim 1.4 \times 10^{20} \cmmoinsdeux$.
The column density has been estimated as
$N_{H} \sim 0.75$--$1    \times 10^{20} \cmmoinsdeux$ \citep{hynes:2000},
$N_{H} \sim 0.75$--$1.15 \times 10^{20} \cmmoinsdeux$ \citep{esin:2001},
$N_{H} \sim 1   $--$1.3  \times 10^{20} \cmmoinsdeux$ 
(McClintock et al. 2001b), 
$N_{H} \sim 0.7$ --$1.5  \times 10^{20} \cmmoinsdeux$ 
(Frontera et al. 2001a), 
all nearly consistent with the mean value in the direction
of the Lockman Hole ($0.5-1.5 \times 10^{20} \cmmoinsdeux$,
see \citealt{hynes:2000}).
From analysis of optical spectra, \citet{dubus:2001} derived 
$N_{H} \sim 1.77$--$4.47 \times 10^{20} \cmmoinsdeux$.
As we see, the exact column density is still a matter of debate,
and will be one of the points discussed in this paper.
McClintock et al. (2001a) 
estimated a distance of $1.8 \pm 0.6 \kpc$,
consistent with the value of $\geq 1.5 \kpc$ derived by \citet{uemura:2000}.

We triggered our multi-epoch multi-wavelength override program
 with {\it HST}/{\it RXTE}/UKIRT to get early observations of 
this system.
We also requested Director's Discretionary
{\it EUVE} observations because of the low value of interstellar
absorption, and obtained the first EUV spectrum of an SXT 
(\citealt{mauche:2000}, \citealt{hynes:2000}). 
We therefore have unprecedented broadband 
coverage, of more than 80\% of the electromagnetic
spectrum from the radio ($\lambda = 21$ cm) to the $\gamma$-rays
(180 keV) (see SED in Fig. \ref{sed_tout}).
The analysis of one of the epochs (corresponding in this
paper to epoch 1, see Fig. \ref{sed_tout_decale}) 
was described in \citet{hynes:2000} 
and we will briefly report here the main results
(see also McClintock et al. 2001b 
and \citealt{esin:2001} for 
an analysis of epoch 2).
The corresponding SED suggested that the system was
exhibiting a low-state mini-outburst, with the inner radius
of the accretion disc estimated to be rather large, 
with a maximum value at $\sim 2000 R_s$ 
($R_s = \frac{2GM}{c^2}$; the Schwarzschild radius for an object of mass $M$) 
demanded by the EUV and X-ray data.
One of the most striking features 
was the strong non-thermal (likely synchrotron)
contribution
in the optical and near-infrared (NIR) wavelengths, characteristic
of outflows seen in jet sources.
Indeed, the SED shows a very flat spectrum from the UV to the
NIR ($\sim 1000$ -- 50\,000 \rm\AA),
suggesting that there is another source of NIR flux dominating any thermal
disc emission, likely related to the radio emission, and therefore
possibly of synchrotron origin.

We followed the evolution of this system from the outburst
towards quiescence, 
to study the mechanisms underlying the outburst.
Preliminary results were published in \citet{chaty:2001e} 
and \citet{chaty:2001d}.
In this paper we will concentrate on the analysis and
evolution of the SED. 
A companion paper \citep{hynes:2003} 
discusses the multiwavelength variability properties.
We describe the multiwavelength observations in Section \ref{observations},
present the results in Section \ref{results},
and discuss them in Section \ref{discussion}.

\section{Observations} \label{observations}

Observational details for our VLA, Ryle telescope, JCMT, UKIRT, 
{\it HST}, {\it EUVE}, {\it SAX}, {\it Chandra} and {\it RXTE}
data follow.

        \subsection{Radio observations}

All the radio (VLA, Ryle Telescope) and sub-mm (JCMT) 
observations are taken from references 
reported in Table \ref{obs_radio}.

\begin{table*}
\begin{flushleft}
\begin{tabular}{|c|c|c|c|c|c|c|c|c|c|} \hline
{ Epoch} & { Date} & { MJD}& { Inst} & { 1.4 GHz} &  { 8.3 GHz} &  { 15 GHz} & { 23 GHz} & { 350 GHz} & { Ref.}   \\ \hline


           & 30--31/03/00    & 51634       & RT  & -           & -           
& 6.2$\pm$0.5               & -           & -   & \citealt{pooley:2000}  \\
           & 31/03/00       & 51635       & RT  & -           & -           
& 7.8$\pm$0.35              & -           & -   & \citealt{dhawan:2000a} \\
\fbox{1}   & 02/04/00       & 51637       & RT  & -           & -           
& 7.5$\pm$0.30              & -           & -   & \citealt{dhawan:2000a} \\
\fbox{1}   & 03/04/00       & 51638       & VLA & 2.1$\pm$0.1 & 6.0$\pm$0.1 
& -                         & $8.8\pm0.3$ & -   & \citealt{dhawan:2000a} \\
           & 05/04/00       & 51640       & RT  & -           & -        
& 8.7$\pm$0.3               & -           & -   & \citealt{fender:2001b}  \\
\fbox{2--6} & 16/03--25/06/00 & 51620--720   & VLA & 2.6$\pm$0.4 & 6.5$\pm$0.7 
& -                         & $9.3\pm1.2$ & -   & \citealt{fender:2001b}  \\
\fbox{2--6} & 16/03--25/06/00 & 51620--720   & RT       & -           & -        
& 9.0$\pm$1.0               & -           &  -       & \citealt{fender:2001b} \\
           & 08--11/00       & 51779--872   & RT       & -           & -        
& 0.15$\pm$0.17             & -           &  -       & Pooley, priv. com.    \\
\fbox{2--6} & 30--31/05/00    & 51695       &JCMT      & -           & -        
& -                         & -           & 41$\pm$4 & \citealt{fender:2001b} \\
           & 09/09/00       & 51796       &JCMT      & -           & -        
& -                         & -           & $<21$    & \citealt{fender:2001b} \\
%
%
\hline
\end{tabular}
\end{flushleft}
\caption[]{\label{obs_radio} { Log of radio observations.} \\
We tabulate here the epoch, date, Modified Julian Date, 
instrument and flux in mJy for every band.
}
\end{table*}

        \subsection{NIR UKIRT observations} \label{nir_ukirt_observations}

Near-infrared observations were carried
out at the UKIRT 3.8 m telescope using UFTI, IRCAM/TUFTI and also CGS4;
some of which were already reported in \citet{hynes:2000}.
The log of the NIR observations is given in Table \ref{obs_infrared}.
The UFTI (UKIRT Fast-Track Imager) 
instrument is a cooled 1--2.5$\mu$m camera with a $1024 \times 1024$ 
pixels$^2$ HgCdTe array. 
The plate scale is 0.091 arcsec per pixel, giving a field of view of 92 arcsec.
The IRCAM/TUFTI instrument is a cooled 1--5$\mu$m camera with a 
$256 \times 256$ pixels$^2$ InSb array.
The plate scale is 0.081 arcsec per pixel with a field of view of 20.8 arcseconds. 

Images were taken through the wide-band filters
J98 ($\lambda  = 1.275 \mu \rm m$; $\Delta \lambda = 0.290 \mu \rm m$), 
H98 ($\lambda  = 1.670 \mu \rm m$; $\Delta \lambda = 0.280 \mu \rm m$) and 
K98 ($\lambda  = 2.205 \mu \rm m$; $\Delta \lambda = 0.41  \mu \rm m$)
with UFTI, and with all the above plus 
L'98 ($\lambda = 3.8   \mu \rm m$; $\Delta \lambda = 0.6   \mu \rm m$) and 
M'98 ($\lambda = 4.675 \mu \rm m$; $\Delta \lambda = 0.250 \mu \rm m$)
with IRCAM/TUFTI.

The exposure times range between 10 s and 60 s.  
The conditions were photometric.  After taking each image of the object,
an image  of the sky  was acquired, to  allow subtraction of the blank
sky. The images were further treated by  removal  of the dark current,
the flat field and  the bright infrared sky.
We also took a NIR K-band spectrum of this source using the
CGS 4 instrument and a $0.6 \asec$ slit
on June, 27.2 UT, which was featureless (see \citealt{chaty:2003}).

\begin{table*}
\begin{flushleft}
\begin{tabular}{|c|c|c|c|c|c|c|c|c|c|c|} \hline
{ Epoch} & { Date}   & { MJD}& { Inst} & { I}        &  { Z}  
&  { J}       &  { H}        &  { K}          & { L'} 
& { M'}       \\ \hline
%

\fbox{1}   & 04/04/00    & 51638.2 & UFTI      &                & 
& 12.12$\pm$0.02 & 11.75$\pm$0.02 & 11.06$\pm$0.02   & 
&                \\
           & 12--15/04/00 & 51647.5 & Sternberg &                & 
& 12.4$\pm$0.2   & 11.9$\pm$0.1   & 10.9$\pm$0.1     &  9.2$\pm$0.1 
&                \\
\fbox{2--4} & 18/04/00    & 51652.5 & TUFTI     &                & 
& 11.92$\pm$0.07 & 11.43$\pm$0.06 & 11.05$\pm$0.08   & 9.71$\pm$0.14 
& 9.38$\pm$0.42  \\
           & 24/06/00    & 51719.2 & UFTI      &                &
&                &                & 11.512$\pm$0.004 & 
&                \\
\fbox{5}   & 26/06/00    & 51721.2 & CGS4      &                &
&                &                & spectrum & 
&                \\
\fbox{6}   & 15/07/00    & 51740.2 & UFTI      &                & 
&                &                & 11.948$\pm$0.006 & 
&                \\
\fbox{7}   & 07/03/01    & 51975   & UFTI      & 17.41$\pm$0.05 &17$\pm$0.05 
& 16.72$\pm$0.05 & 16.15$\pm$0.05 & 15.77$\pm$0.05   & 
&                \\
%
\hline
\end{tabular}
\end{flushleft}
\caption[]{\label{obs_infrared} { Log of the infrared observations.} \\
The epoch, date, MJD, instrument and magnitudes for every filter are
reported.
The magnitudes of the first epoch (corresponding to the detection
of the infrared counterpart) were reported in \citet{chaty:2000},
and those from Sternberg in \citet{taranova:2000}.
The observations of epochs 5 and 7 are reported in more details
in \citet{chaty:2003}.
}
\end{table*}

        \subsection{Optical and Ultraviolet {\it HST} observations}

{\it Hubble Space Telescope} ({\it HST}) observations were performed with
the Space Telescope Imaging Spectrograph (STIS; \citealt{leitherer:2001}) on
the dates indicated in Table~\ref{obs_optical-uv}.  
These spanned the UV and optical bands 
at high and low resolution respectively 
using the E140M, E230M, G430L and G750L modes.  
For each visit, average calibrated spectra were extracted from standard
{\it HST} pipeline data products.  There was useful coverage from
1150--10000\,\AA, although the region from 1195--1260\,\AA\ was completely
dominated by Ly$\alpha$ absorption and N\,{\sc v} emission 
\citep{haswell:2002}, so was excluded from our spectral energy distributions.
On the dates where more than one observation of the same wavelength range
was taken, the
spectra were averaged to increase the signal-to-noise ratio.
The documented absolute calibration accuracy is 5\% for the optical (CCD) 
modes, and 8\% for the UV (MAMA) modes.  The break between them is around 
3100\AA, with some overlap.  This systematic uncertainty is larger 
than any statistical uncertainties.

\begin{table*}
\begin{flushleft}
\begin{tabular}{|c|c|c|c|c|c|c|c|} \hline
{ Epoch}&{ Date}& { MJD}&{ E140M}&\multicolumn{2}{c}{E230M} & { G430L}&{ G750L}\\ \hline
\multicolumn{3}{c}{Central wavelength (\AA):} & { 1425}&{ 1978}&{ 2707}&{ 4300}&{ 7751} \\ \hline
\multicolumn{3}{c}{Wavelength range (\AA):} &{ 1123--1710}&{1574--2382}&{2303--3111}&{ 2900--5700}&{5236--10266} \\ \hline


\fbox{1} & 08/04/00 & 51642.7 & 9150     & 1300    & 1200    & 144    & 180 \\
%
\fbox{2} & 18/04/00 & 51652.6 & 3000     & 1000    & 700     & 1160   & 150 \\
\fbox{3} & 28/04/00 & 51663.3 & 1620     & 820     & 700     & 120    & 150 \\
\fbox{4} & 28/05/00 & 51692.8 & 1800     & 1000    & 750     & 120    & 204 \\
\fbox{5} & 24/06/00 & 51720.0 & 1750     & 1000    & 700     & 158    & 216 \\
         &          &         & 3000     &         &         &        &     \\
\fbox{6} & 08/07/00 & 51733.9 & 1700     & 950     & 750     & 120    & 224 \\
\fbox{7} & 28/09/00 & 51815   & 1250     & 2050    & 2050    & 400    & 407 \\
%
\hline
\end{tabular}
\end{flushleft}
\caption[]{\label{obs_optical-uv} { Log of the optical and UV observations.} \\
The epoch, date, MJD, central wavelength and wavelength range are indicated, 
with in each case the exposure time in seconds.
}
\end{table*}

        \subsection{Extreme Ultraviolet {\it EUVE} observations}

{\it Extreme Ultraviolet Explorer\/} ({\it EUVE\/}) observations of
$\xtejodh$ took place during 2000 April 8.10--8.71, 13.32--13.93,
and 16.91--19.60 UT.  They are described in \citet{hynes:2000}.
The log of the {\it EUVE} observations is reported in Table \ref{obs_euve}.
As in \citet{hynes:2000}, we considered only the data between 70
and 120 \AA, since longer wavelengths are heavily absorbed and dominated
by noise. 
In the following plots showing data from {\it EUVE}, 
the vertical error bars are the 1 $\sigma$ errors from the photon statistics.

\begin{table}
\begin{flushleft}
\begin{tabular}{|c|c|c|c|} \hline
{ Epoch}& { Date}  & { MJD}            & { Exp. time} \\ \hline


\fbox{1} & 8--9/04/00   & 51642.603--51643.213 & 19193.52         \\
         & 13--14/04/00 & 51647.821--51648.430 & 19553.9          \\
\fbox{2} & 16--19/04/00 & 51651.407--51654.102 & 80200.1          \\
%
\hline
\end{tabular}
\end{flushleft}
\caption[]{\label{obs_euve} { Log of the {\it EUVE} observations.} \\
The epoch, date, MJD, and exposure time in seconds are indicated.
}
\end{table}

        \subsection{X-ray {\it SAX} observations}

Beppo-{\it SAX} observed $\xtejodh$ 4 times, on 2000 April 14, May 4, June 26
and December 12. Here we show the 
observation
taken on 2000 April 14--15, and reported in Table \ref{obs_sax}.
The details of this observation can be found in 
Frontera et al. (2001a). 
Since these data were initially corrected with an interstellar absorption of 
$1.5 \times 10^{20} \cmmoinsdeux$, we first uncorrected these data
to get the observed flux, and then corrected them from the
absorption with the desired value
of $N_H$, as described in section \ref{correction}.

\begin{figure}
\centerline{\psfig{file=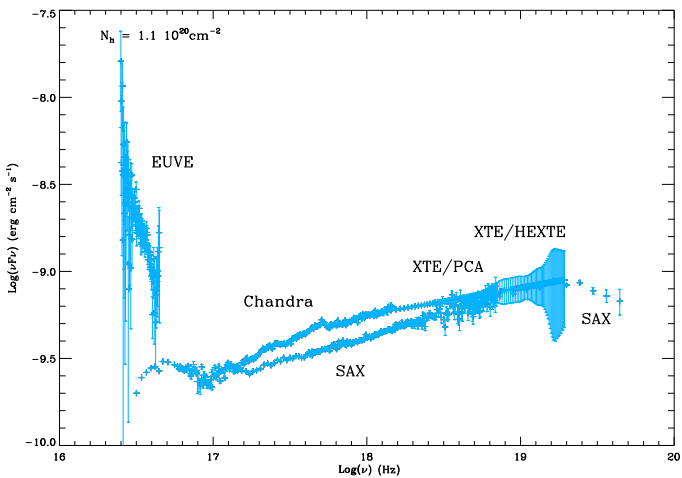,angle=90.,width=9.cm}}
\caption[]{\label{sed_vis2_calib} 
Comparison of epoch 2 Spectral Energy Distribution as
observed by the instruments {\it EUVE}, {\it Chandra}, {\it SAX} and {\it XTE}.
The SED is corrected with 
$N_H = 1.10 \times 10^{20} \cmmoinsdeux$.}
\end{figure}

As shown in Figure \ref{sed_vis2_calib}, 
although there is a good agreement between
{\it SAX} and {\it Chandra} data in the interval $\log \nu = 16.6$--$17.1$
and between {\it SAX} and {\it RXTE} data for the interval
$\log \nu = 18.5$--$18.7$, 
there is an inconsistency between the {\it SAX} and the {\it EUVE} 
observations for the interval $\log \nu = 16.5$--$16.7$
(likely due to the model used in fitting the {\it SAX} data
combined with the low response of the detector at those wavelengths) 
and in the interval $\log \nu = 17.0$--$18.5$ 
between {\it SAX} and both {\it RXTE} and {\it Chandra} observations. 
This seems to be a calibration problem.

Nonetheless, the SAX spectrum unambiguously shows the presence of a cut-off
at an energy of $\log \nu \sim 19.5$, 
as discussed by Frontera et al. (2001a). 
Therefore in the following we will include the SAX data only at high energies
(in the interval $\log \nu = 18.8$--$19.65$) 

\begin{table*}
\begin{flushleft}
\begin{tabular}{|c|c|c|c|c|c|c|} \hline
{ Epoch} & { Date} & { MJD} & { LECS}& { MECS}& { HPGSPC}& { PDS} 
\\ \hline
\multicolumn{3}{c}{Energy bands (keV):}&{ 0.12--4}&{ 1.7--10}&{ 7--29}&{ 15--200} \\ \hline
%
%
\fbox{2}  & 14--15/04/00 & 51648--9 & 21197 & 29758 & 42138 & 20146 \\
%
\hline
\end{tabular}
\end{flushleft}
\caption[]{\label{obs_sax} { Log of the {\it SAX} observations.} \\
The epoch, date, MJD, energy bands and exposure time in seconds 
for the different instruments are indicated.
For more details about the {\it SAX} observations see 
Frontera et al. (2001a). 
}
\end{table*}

        \subsection{X-ray {\it Chandra} observations}

There was only one observation by {\it Chandra}, taken on 2000,
April 18, reported in Table \ref{obs_chandra}.
Details are in McClintock et al. (2001b). 
Since these data were corrected with an interstellar absorption of 
$1.3 \times 10^{20} \cmmoinsdeux$, as for the {\it SAX} data,
we first uncorrected these data
to get the observed flux, and then corrected them for the
absorption with the desired value
of $N_H$, as described in paragraph \ref{correction}.

\begin{table}
\begin{flushleft}
\begin{tabular}{|c|c|c|c|} \hline
{ Epoch}& { Date}  & { MJD}            & { Exp. time} \\ \hline

%
\fbox{2} & 18/04/00    & 51652.5             & 27200          \\
%
\hline
\end{tabular}
\end{flushleft}
\caption[]{\label{obs_chandra} { Log of the {\it Chandra} observations.} \\
The epoch, date, MJD and exposure time in seconds are indicated.
For more details about the {\it Chandra} observations see 
McClintock et al. (2001b). 
}
\end{table}

        \subsection{X-ray {\it RXTE} observations and data analysis}

We observed $\xtejodh$ with the {\it Rossi X-ray Timing Explorer (RXTE)}
Proportional Counter Array (PCA) and High-Energy Timing Experiment
(HEXTE) at five epochs selected to coincide with the {\it HST}
visits. The log of the {\it RXTE} 
observations is reported in Table \ref{obs_rxte}.
The method used for the analysis of data is the same 
than the one described in \citet{hynes:2000},
therefore we will just give in the following the
various parameters derived by analysing the different epochs.

For the PCA, we used the ``standard mode'' data 
(128 spectral channels, 16-second
accumulations), selecting sub-intervals when the number of
detectors on remained constant (about 90\% of the total time). We
similarly extracted 256-channel spectral accumulations from the
HEXTE science event (SE) data. A subset of PCA and HEXTE
detector channels, corresponding typically to about 3-100 keV were
used in our subsequent model fitting. Background rates for the PCA
were estimated using the epoch-4 models, and response matrices
were generated using the current calibration files and
response-matrix generation software, all from the ``HEAsoft 5.1''
release.

The source intensity was typically in the 30-40
mCrab range for each epoch, with typical PCA count rates of 
$\sim 80-120$ cts/sec/PCU (source; the background is an additional 30/s/PCU).
The spectra thus
derived were found to be hard, with photon power-law indices of
about $1.8\pm0.1$. A thermal Comptonization model \citep{sunyaev:1980}
with $\tau\simeq 3$ and $T_{\rm e}\simeq
30$\,keV, also provided acceptable fits (in either the power-law or thermal
Comptonization cases, typical $\chi^2$ per degree of freedom of order unity
were obtained). 
There was no evidence (in terms
of statistical improvement to our fits) for a soft-excess component,
thus we conclude, as have others, that the source remained in the
``low/hard'' spectral state throughout the outburst.
In most cases, particularly 50133-01-03-00 for which the PCA exposure 
was about 10,000 seconds, there was a distinct positive residual 
corresponding to the 6.4\,keV FeK resonance, thus a Gaussian  line 
profile was included to refine the overall fit.
The energy coverage was $\sim 3$ to $26$ keV with PCA
and $11$ to $207$ keV with HEXTE.
We show in all the Figures only the data up to $80$ keV
(except for the first visit where we show them up to $120 \keV$),
since at higher energies the noise dominates.

\begin{table}
\begin{flushleft}
\begin{tabular}{|c|c|c|c|c|} \hline
{ Epoch}& { Date}& { MJD} & Exp. time & Observation ID \\ \hline

%
\fbox{1} & 08/04/00 & 51642.7 &  3900 & 50133-01-01-00           \\
\fbox{2} & 18/04/00 & 51652.6 &  4700 & 50133-01-02-00          \\
\fbox{3} & 28/04/00 & 51663.3 & 10600 & 50133-01-03-00           \\
\fbox{4} & 28/05/00 & 51692.8 &  2400 & 50133-01-04-00         \\
\fbox{5} & 24/06/00 & 51720.0 &  5600 & 50133-01-05-00        \\
%
\hline
\end{tabular}
\end{flushleft}
\caption[]{\label{obs_rxte} { Log of the {\it RXTE} observations.} \\
We report here the epoch, date, MJD, exposure time in seconds 
for the PCA instrument and the observation ID.
}
\end{table}

        \subsection{Broadband SED}

\begin{figure}
\centerline{\psfig{file=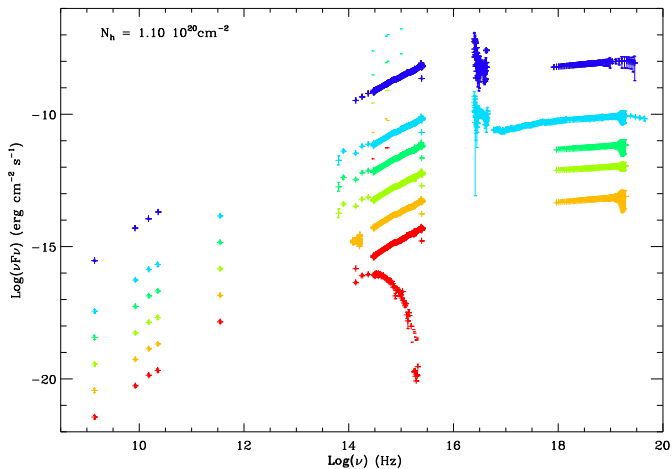,angle=90.,width=9.cm}}
\caption[]{\label{sed_tout_decale} 
{ Spectral Energy Distributions of all the epochs, beginning from the top
of the Figure. Colour-codings are as given in Table \ref{table_epochs}. 
For easier reading, we multiplied epoch 1 by 10, and then
divided epoch 2 by 10, epoch 3 by 100, epoch 4 by $10^3$,
           epoch 5 and 7 by $10^4$ and epoch 6 by $10^5$.}
}
\end{figure}

\begin{figure}
\centerline{\psfig{file=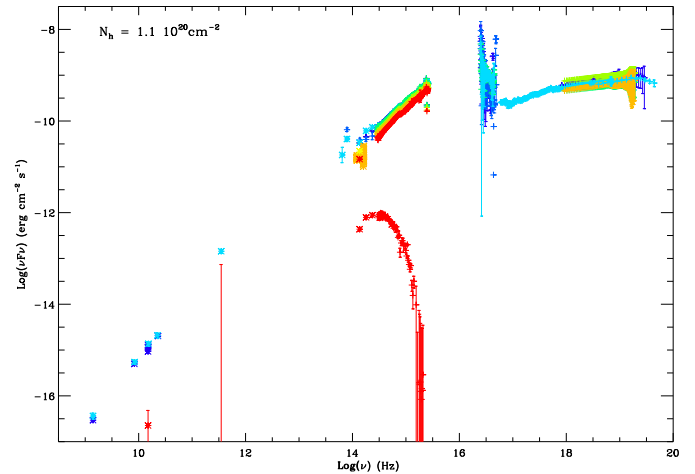,angle=90.,width=9.cm}}
\caption[]{\label{sed_tout} { Spectral Energy Distribution of all the epochs
overplotted}
}
\end{figure}

\begin{figure}
\centerline{\psfig{file=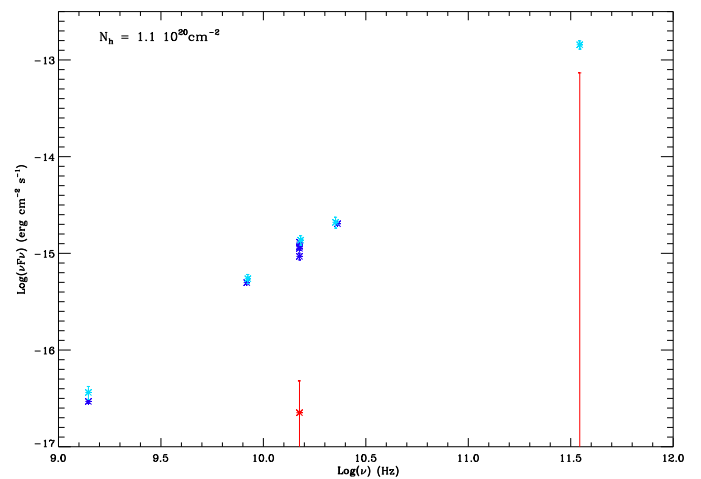,angle=90.,width=9.cm}}
\caption[]{\label{sed_tout_radio} { Radio Spectral Energy Distribution 
of all the epochs}
}
\end{figure}

\begin{figure}
\centerline{\psfig{file=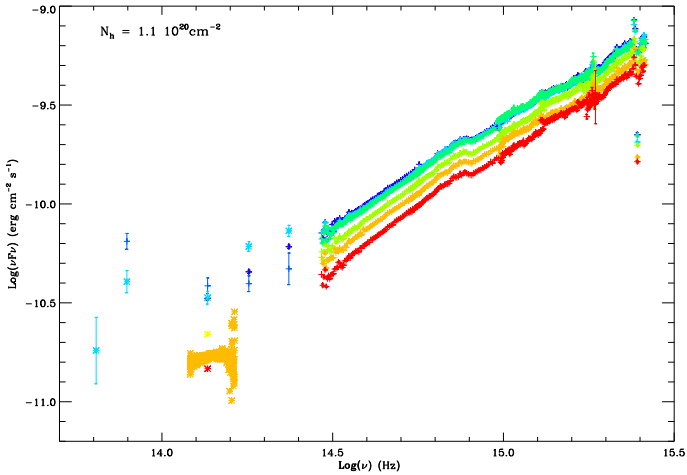,angle=90.,width=9.cm}}
\caption[]{\label{sed_tout_nir-uv} { NIR-UV Spectral Energy Distribution 
of all the epochs}
}
\end{figure}

\begin{figure}
\centerline{\psfig{file=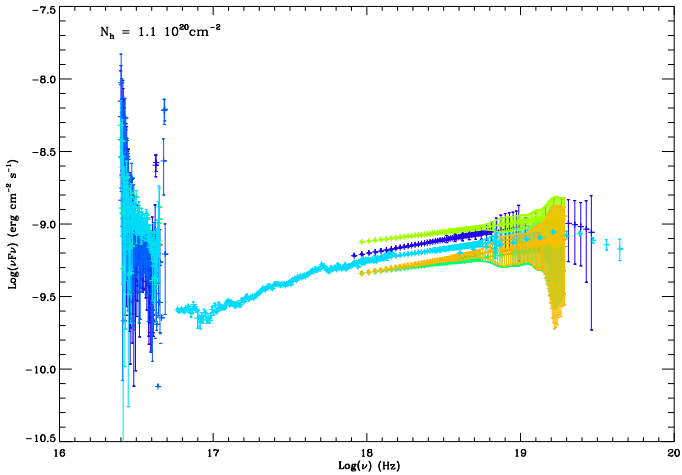,angle=90.,width=9.cm}}
\caption[]{\label{sed_tout_euv-x} { EUV-X Spectral Energy Distribution 
of all the epochs}
}
\end{figure}

For analysis, we separated the observations
into the 7 epochs reported in Tables \ref{obs_radio}--\ref{obs_rxte}.
Table \ref{table_epochs} summarizes the different
facilities used at each epoch.
For each epoch, observations in all bands are simultaneous
or nearly-simultaneous.
The broadband SEDs corresponding to all epochs are presented
in Figures \ref{sed_tout_decale} and \ref{sed_tout}.
In Figure \ref{sed_tout_decale} they are shown 
with different normalisations for more clarity.
We overplot all epochs in Figure \ref{sed_tout},
and enlargements in radio, NIR--UV and EUV--X-ray regions are
shown respectively in 
Figures \ref{sed_tout_radio}, \ref{sed_tout_nir-uv} and \ref{sed_tout_euv-x}.
In all these figures the data have been corrected with 
$N_H = 1.1 \times 10^{20} \cmmoinsdeux$ corresponding to $A_v = 0.059$, 
as will be described in section \ref{correction}.

\begin{table*}
\begin{flushleft}
\begin{tabular}{|c|c|c|c|c|c|c|c|c|c|c|} \hline
{ Epoch}& { Date}   & { MJD} & { Radio} & { UKIRT} & { HST} &
{ EUVE} & { SAX} & { Chandra} & { XTE} & { colour} \\ \hline
\multicolumn{3}{c}{$\log(\nu)$:} & {\scriptsize  9.0--11.6} & 
{\scriptsize  13.78--14.48} & {\scriptsize  14.47--15.41} & 
{\scriptsize  16.38--16.61} & {\scriptsize  16.61--19.68} & 
{\scriptsize  16.76--18.23} & {\scriptsize  17.78--19.68} & \\ \hline
%
%
\fbox{1} & 08/04/00 & 51642.7 & x & x & x & x &   &   & x & dark blue   \\
\fbox{2} & 18/04/00 & 51652.6 & x & x & x & x & x & x & x & light blue  \\
\fbox{3} & 28/04/00 & 51663.3 & - & - & x &   &   &   & x & dark green \\
\fbox{4} & 28/05/00 & 51692.8 & - & - & x &   &   &   & x & light green  \\
\fbox{5} & 24/06/00 & 51720.0 & - & x & x &   &   &   & x & yellow      \\
\fbox{6} & 08/07/00 & 51733.9 & - & x & x &   &   &   &   & orange      \\
\fbox{7} & 09/00--03/01 & 51815--52467 &  & x & x &   &   &   &   & red         \\
%
\hline
\end{tabular}
\end{flushleft}
\caption[]{\label{table_epochs} { Log of the different epochs.} \\
The date and MJD are indicated. The ``-'' indicates that we used
for the indicated epochs the observations of epoch 2,
which is sensible since the SED did not change significantly in these
wavelengths during 3 months.
}
\end{table*}






\section{Results} \label{results}

Before describing the results, we present in Section \ref{correction}
the method used to correct for interstellar absorption.

        \subsection{Correction of interstellar absorption} \label{correction}

The interstellar absorption was corrected in the following way.
First, we choose a value for the column density 
by fitting the UV and EUV fluxes and slopes,
as will be discussed in section \ref{nh}.
For example here we take $N_H = 1 \times 10^{20} \cmmoinsdeux$.

Using this value of column density, 
we then correct the {\it EUVE} data and {\it SAX} and {\it Chandra} X-ray data
 using the absorption 
cross sections of \citet{rumph:1994} for HI, HeI and HeII with abundance
ratios 1:0.1:0.01, typical of the diffuse interstellar medium.

Then, we assume that this inferred column density is interstellar,
and we adopt an average gas-dust ratio of 
$\langle N(H_I+H_2)/E(B-V) \rangle = 5.8 \times 10^{21} \cmmoinsdeux \mag^{-1}$
\citep{bohlin:1978}.
This leads to the value of $E(B-V)$, equal to $0.017$ 
in the example taken here.
Taking the value of $R_v \equiv A_v / E(B-V) = 3.1$ 
typical of the diffuse interstellar medium \citep{cardelli:1989},
we get the value of $A_v$ ($=0.053$ here).

Finally, we correct the infrared (from UKIRT), 
optical and ultraviolet ({\it HST})
observations for interstellar absorption 
with the extinction law of \citet{cardelli:1989},
using the derived value of absorption, leading to the inferred SED.

We note that even in the FUV, reddening corrections are $\lesssim$ 15 \%
(depending on E(B-V)) and that extinction curve and gas-to-dust scaling
are not critical. Only in the EUV is the correction large.

	\subsection{Geometrical parameters of the system}

We took a black hole mass of $7.2 \pm 1.3 \Msol$ (McClintock et al. 2001a),
which corresponds to a Schwarzschild radius of $R_s = 21$ km.
The mass ratio has been measured as $Q = 1/q = \frac{M_1}{M_2} = 27 \pm 5$
 (an extreme value among SXTs, \citealt{orosz:2001}), therefore
the mass of the donor star $M_2 = 0.27 \pm 0.05 \Msol$.
The distance of the system has been determined as $1.71 \pm 0.05 \kpc$ 
(see discussion in Section \ref{quiescent}) 
and the orbital period to $P_{\rm orb} = 0.169937(1)$ 
days $\sim 4.08$ hours \citep{zurita:2002}.

This gives an orbital separation 
$a = 1.76 \pm 0.1 \times 10^9$ m \citep{paczynski:1971}.
To derive the value of the outer radius of the accretion disc,
we take the intermediate value between the disc's tidal radius
$R_T$ and the 3:2 resonant radius $R_{23}$, since the source
showed the presence of superhumps (see Section \ref{introduction}).
The disc's tidal radius is taken as 90\% of the Roche lobe radius,
therefore 
$R_T/a = 0.58 \pm 0.01$ \citep{eggleton:1983}
and the 3:2 resonant radius is $R_{23}/a \sim 0.47$ \citep{whitehurst:1991}.
Hence the outer radius we take is $r_{out} = 0.52a$.
The inclination of the system is chosen to be $70 \pm 10 \deg$, 
consistent with McClintock et al. (2001a, 
and also \citealt{zurita:2002}).

The inner radius of the accretion disc will be a free parameter,
but will typically be between 300 and 450 $R_s$. This corresponds
to a low state, in contrast to the high state where the accretion
disc extends very close to the compact object, $r_{in} \approx 3 \times R_s$,
corresponding to the last stable orbit.

        \subsection{Fit to the nearly-quiescent SED} \label{quiescent}

\begin{figure}
\centerline{\psfig{file=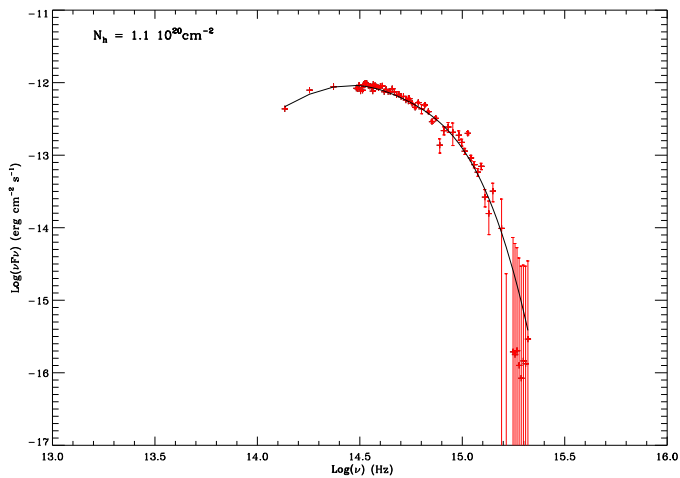,angle=90.,width=9.cm}}
\caption[]{\label{sed_fit_vis7} {
Fit of the epoch 7 (the nearly-quiescent system), with 
a fractional contribution secondary/total emission of $\sim 25 \%$,
a remnant accretion disc at $\sim 6000 K$
and a distance of $\sim 1.7 \kpc$, taking into account
the emission of a M1V star at $T_{eff} = 3400$K with a radius
$R_2 = 0.39 \Rsol$.
}
}
\end{figure}

To characterize the nearly-quiescent system,
we use only the epoch 7 data from UKIRT and {\it HST}.
Since there is probably still some contamination from the 
accretion disc, we take the combination of two emission models, one
representing the companion star and one representing 
the accretion disc. 
The spectral type of the companion star has been determined as
K5--M1 V (McClintock et al. 2001a, 
\citealt{wagner:2001}).
In our analysis, we therefore 
take an M1 V star photosphere ($T_{eff} = 3400 K$, \citealt{bessell:1998}).
The radius of the mass donor star can be estimated with 
$R_2/a = 0.15 \pm 0.01$ \citep{eggleton:1983} therefore $R_2 = 0.39 \Rsol$.
We take a black body model for the nearly-quiescent accretion disc.
We therefore fit the epoch 7 UKIRT and {\it HST} data with 
 three free parameters:
the fractional contribution of the secondary to the total emission,
the temperature of the accretion disc,
and the distance of the system.
We found that the best-fit parameters were 
a fractional contribution of $25 \pm 2 \%$,
a remnant accretion disc at $6000 \pm 50 K$
and a distance of $1.71 \pm 0.05 \kpc$ with a reduced $\chi^2 = 1.3$, 
and a tolerance of $10^{-5}$
(see 
Figure \ref{sed_fit_vis7}).
The fractional contribution of the secondary to the total emission 
which we derived 
is in agreement with the contribution
of $\sim 28-36 \%$ found by \cite{wagner:2001} 
and $\sim 34 \pm 8\%$ by McClintock et al. (2001a) 
from optical
spectroscopy.
We also tried to replace the black body model by a power-law, 
but the fit was worse, showing that the synchrotron contribution
had decreased between epochs 1-6 and 7.
We finally varied the absorption through
the column density interval allowed by the broad band SED 
(see section \ref{nh})
but this did not change substantially the best fit parameters.
Therefore our simple model gives a result consistent with the
other methods, and
in the following we will use the parameters derived from our analysis, 
to add the contribution from the companion star to the overall fits
at other epochs.

It is interesting to point out that, 
by plotting the infrared colour-magnitudes
(observed on 2001, March, during the near-quiescence of the source,
corresponding to epoch 7) on a Hertzsprung-Russell diagram, 
we derived that the companion
star had a spectral type of M1 V \citep{chaty:2003}, in
agreement with the spectroscopic results.
On the other hand, 
the radius of the companion star that we used for the fit in Figure \ref{sed_fit_vis7}
is smaller than an
 isolated main sequence star by a factor two. However,
\cite{haswell:2002} showed that $\xtejodh$'s companion star is an evolved star,
consequently it may well have a different radius to an isolated
main sequence star of the same type.

        \subsection{Epoch 2} \label{fit_epoch2}

As we can see in Fig. \ref{sed_tout} 
(which we will discuss in more detail in section \ref{fit_different}),
there were no gross changes in the SED between epochs 1 and 6, so in the
following we will consider the epoch where we have the most data,
i.e. epoch 2. 
Our analysis differs from previous ones (see Section \ref{introduction})
because, in addition to the disc black body, 
we explicitely include contributions in different bands
(for instance power-law spectra in radio, submm-NIR and X-rays), 
together with the emission in the NIR-optical domain
of the nearly-quiescent system, using the results from 
Section \ref{quiescent}.
We are aware that this does not account for irradiation of the mass
donor star, however
the fits are just intended to be illustrative.
$\xtejodh$ is a unique source in term of its low absorption,
allowing us to study its whole SED in great detail,
derive some useful constraints, and
analyse the evolution from the outburst to the near-quiescence.

                 \subsubsection{Model of the accretion disk} \label{accretion}

To fit the broad band emission due to the accretion disc we use
the simple parameterised model of \citet{hynes:2002a}.
It is based on a combination of the classic viscously heated black
body disc spectrum (\citealt{shakura:1973}, \citealt{frank:1992})
and the modified temperature distribution for an irradiated disc
(\citealt{cunningham:1976}, \citealt{vrtilek:1990}).
See these papers for derivations of the relevant temperature
distributions, and \citet{dubus:1999} for a critique of the assumptions.

The model spectrum is calculated by summing a series of black
bodies over radius. The local effective temperature of a disc
annulus is determined by the emergent flux at that radius,
such that $T_{\rm eff}^4 \propto F_{\rm bol}$.
The emergent flux is the sum of viscous energy release within that
annulus, $F_{\rm visc} \propto T_{\rm visc}^4$, and the X-rays
reprocessed by the annulus, $F_{\rm irr} \propto T_{\rm irr}^4$.
Hence the effective temperature contains contributions 
from viscous heating ($T \propto R^{-3/4}$) and 
irradiation ($T \propto R^{-3/7}$).
The effective temperature profile of the disc is therefore represented by
\[ T_{\rm eff}^4(R) = T_{\rm visc}^4(R) + T_{\rm irr}^4(R)\,. \]

Both profiles are effectively controlled in the model
by the temperature at the outer
radius of the disc. The viscous temperature, $T_{visc}(R_{\rm out})$, 
is a free parameter, typically around 7500 K; 
the irradiation temperature, $T_{irr}(R_{\rm out})$, we usually take to be 
$T_{\rm irr}(R_{\rm out}) = 0 $ K (but see Section \ref{irradiation}).

The detailed fit is the sum of the secondary star
(or nearly-quiescent system),
the model of the accretion disc in the low state, and
three different power-laws demanded by the data
in the radio, NIR, and X-ray bands,
with respective spectral indices $0.5$, $-0.15$ and $-0.8$, taking the
convention 
\begin{equation} \label{spectral}
f_\nu \propto \nu^\alpha.
\end{equation}

As the figures show, these three power-laws are natural fits
to the SED.
They have the respective expressions: 
${f_\nu}_{\rm radio} =   7 \times 10^{-31} \times \nu^{0.5} $ 
between $1 \times 10^9$  and $4 \times 10^{11}$ Hz; 
${f_\nu}_{\rm nir}   = 2.5 \times 10^{-23}\times\nu^{-0.15}$ 
between $4 \times 10^{11}$ and $8 \times 10^{14}$ Hz; and
${f_\nu}_{\rm x}     = 1.2 \times 10^{-13} \times \nu^{-0.8}$ 
between $8 \times 10^{14}$ and $10^{19.3}$ Hz.


                   \subsubsection{Examining the SED}

\begin{figure}
\centerline{\psfig{file=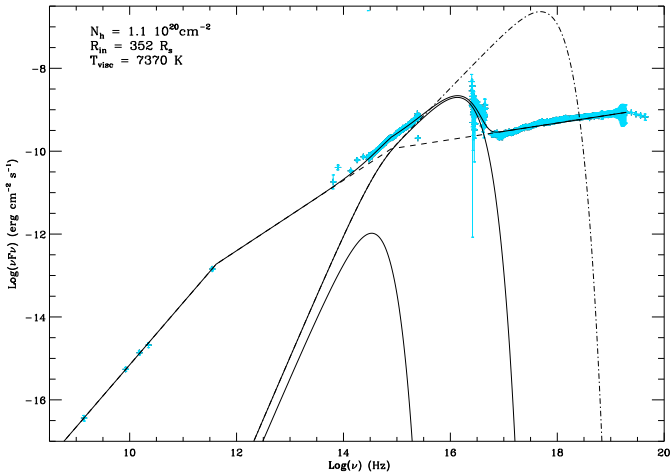,angle=90.,width=9.cm}}
\caption[]{\label{sed_vis2+mod} 
{ Spectral Energy Distribution of epoch 2 corrected with 
$N_H = 1.10 \times 10^{20} \cmmoinsdeux$.}
$R_{\rm in}$ and $T_{visc}$ given in the Figure are those derived by our fits.
The lower solid thick curve is the emission representing the near-quiescent
system, 
the intermediate solid thin curve is the multicolour black body disc 
in the low state,
and the upper dash dot thin curve 
is the multicolour black body disc in the high state.
The straight dashed lines are the three different power-laws, as described in
section \ref{fit_epoch2}.
The solid curve following the data is the sum of the contributions
from the near-quiescent system, the accretion disc in the low state, 
and the three power laws.
} 
\end{figure}

We show epoch 2 with the different models and fit 
in Figure \ref{sed_vis2+mod}. We see clearly two main characteristics.
Firstly the source is exhibiting a very low low/hard state \citep{hynes:2000}; 
secondly there is a strong non-thermal contribution in the
radio domain with an inverted spectrum, extending up to the UV wavelengths, 
and even in the X-rays. We will develop this later.

In most of our fits the typical values of the inner disc temperature,
$T_{\rm in}$ are between 20 and 30 eV and those of the inner disc radius
 $R_{\rm in}$ are between $300$ and $450 R_{\rm s}$ 
(see Section \ref{nh} below and Table \ref{table_nh}).
These values are consistent with those derived by McClintock et al. (2001b)
but the inner radii are higher than that derived by \citet{esin:2001},
consistent with their model including a significant ADAF contribution
to the EUV.

        \subsection{The hydrogen column density, $\nh$} \label{nh}

We now try to better constrain the value of $\nh$ by fitting 
the whole SED corrected with different values of $\nh$, with the
inner radius of the accretion disc and the outer temperature
as free parameters. All the other parameters are taken as
described in previous sections. 
The results are reported in Table \ref{table_nh}.
The best fits had a reduced $\chi^2$ of 54, with 728
degrees of freedom. 
This high value of the reduced $\chi^2$
just shows how illustrative the fits are, 
because detailed spectral features such as the Balmer jump 
and the dip in the Chandra spectrum are not fitted. 
From the $\chi^2$ values obtained and inspection of corresponding 
figures, reasonable fits appear to be with $\nh$ between $0.8$ and 
$1.3 \times 10^{20} \cmmoinsdeux$. 

With $\nh=1.45 \times 10^{20} \cmmoinsdeux$ 
($R_{\rm in} = 313 R_s$) 
no accretion disc models are simultaneously consistent with 
the very different slopes in the UV and the EUV.
With $\nh = 0.75 \times 10^{20}$ 
($R_{\rm in} = 970 R_s$) 
the presence of the X-ray power law
makes it impossible to reconcile both high UV and low EUV fluxes.
$\nh = 1.1 \times 10^{20} \cmmoinsdeux$ is the only one of the values 
considered here which is consistent with both accretion
disc models and the UV/EUV fluxes and slopes.
We present the epoch 2 SED corrected with the 
column density $\nh = 1.1 \times 10^{20} \cmmoinsdeux$ 
in Figure \ref{sed_vis2+mod}.

Therefore, from the results of the fits, and also by plotting these
 results on the observations,
$\nh = 0.8-1.3 \times 10^{20} \cmmoinsdeux$ seems to be the preferred value.
In the following we will consider the value of
$\nh \sim 1.1 \times 10^{20} \cmmoinsdeux$  as the most likely,
and we will draw all the figures with this value. 

\begin{table}
\begin{flushleft}
\begin{tabular}{|c|c|c|c|c|} \hline
{ $\nh$ ($\times 10^{20}$)} & { $R_{in}$ ($R_s$)} & { $T_{visc}$ (K)} &
{ $T_{in}$ (eV)} & { $\chi^2$} \\ \hline

%
%
%
%
%
0.70*                       & 1369           & 7373 & 8.53 &  65          \\
0.75*                       & 970           & 7320 & 10.97 &  58          \\
0.80                       & 450           & 7246 & 19.32 &  54          \\
0.85                       & 427           & 7266 & 20.15 &  54          \\
0.90                       & 407           & 7286 & 20.94 &  54          \\
0.95                       & 389           & 7306 & 21.72 &  55          \\
1.00                       & 374           & 7327 & 22.44 &  55          \\
1.05                       & 362           & 7348 & 23.06 &  56          \\
{\bf 1.10}             & {\bf 352}  & {\bf 7370} & {\bf 23.62} &  {\bf 56}     \\
1.15                       & 354           & 7394 & 23.60 &  57          \\
1.20                       & 354           & 7418 & 23.67 &  57          \\
1.25                       & 356           & 7442 & 23.65 &  58          \\
1.30                       & 326           & 7460 & 25.27 &  58          \\
1.35                       & 310           & 7480 & 26.37 &  59          \\
1.40                       & 316           & 7505 & 26.08 &  60          \\
1.45                       & 313           & 7528 & 26.35 &  60          \\
1.50                       & 310           & 7551 & 26.62 &  61          \\
1.55                       & 307           & 7575 & 26.90 &  62          \\
1.60                       & 308           & 7599 & 26.92 &  62          \\
1.65                       & 306           & 7622 & 27.13 &  63          \\
1.75                       & 302           & 7669 & 27.57 &  65          \\
1.90                       & 296           & 7739 & 28.25 &  67          \\
2.05                       & 292           & 7810 & 28.80 &  70          \\
%
\hline
\end{tabular}
\end{flushleft}
\caption[]{\label{table_nh} Results of different fits according to $\nh$.
The $\nh$ is fixed, $R_{\rm in}$ and $T_{\rm visc}$ 
are free parameters. Although it
is not an independent parameter in the fits, we give $T_{\rm in}$ for information.
The tolerance used in the fits was $1 \times 10^{-2}$ (except where
an asterisk is written, where the tolerance was between 0.1 and 1).
Reasonable fits appear to be with $\nh$ between $0.8$ and 
$1.3 \times 10^{20} \cmmoinsdeux$, and by inspecting the figures
$1.1 \times 10^{20} \cmmoinsdeux$ seems to be the best value,
and we will keep it for the rest of the paper
(see text for a discussion of this).
}
\end{table}




        \subsection{Irradiation} \label{irradiation}

\begin{figure}
\centerline{\psfig{file=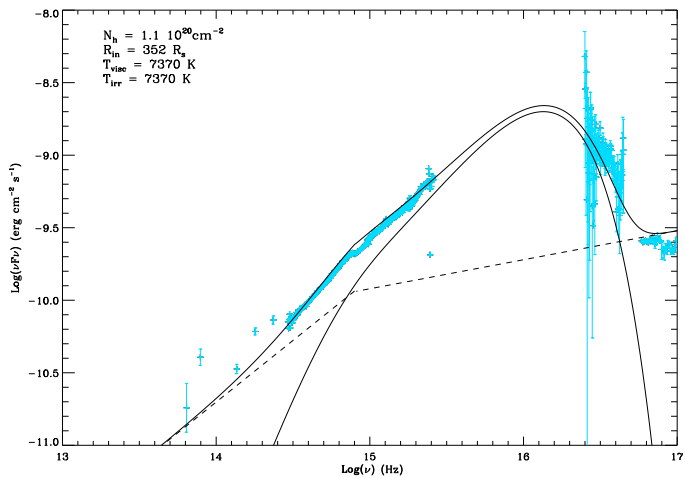,angle=90.,width=9.cm}}
\caption[]{\label{sed_vis2_ir8k} 
{ Spectral Energy Distribution of epoch 2 corrected with 
$N_H = 1.1 \times 10^{20} \cmmoinsdeux$, $R_{\rm in} = 352 R_s$
and with an irradiation of $T_{\rm irr}(R_{\rm out}) = 7370 $ K 
(see section \ref{accretion}).}
} 
\end{figure}

In Figure \ref{sed_vis2_ir8k} we present 
the epoch 2 SED corrected with the
column densities $1.10 \times 10^{20} \cmmoinsdeux$, 
taking into account an irradiation 
with $T_{\rm irr}(R_{\rm out}) = 7370 $ K, 
as described in section \ref{accretion}.
Figure \ref{sed_vis2+mod} shows the same SED
without any irradiation.
We can see that the slopes in the optical/UV parts of the spectrum
are not consistent with the presence of irradiation.
However, the irradiation component in our fit 
presents lots of assumptions, mainly about the geometry of
 the irradiating/irradiated region. 
Therefore, in view of the uncertainties in the irradiation function,
although it is not required by our characterization of the data
(which is also very crude),
some irradiation of the disc cannot be ruled out.

        \subsection{Contributions to $L_{\rm Bol}$} \label{contributions}


The luminosity due to thermal-viscous dissipation
in the accretion disc is given by integrating the luminosity
$\sigma T^4$ between $R_{in}$ and $R_{out}$, for the two faces of the disc:

\[L_{\rm disc} = 2 \times 2 \pi \sigma T_{\rm out}^4 \times R_{\rm out}^{3} 
\times (\frac{1}{R_{\rm in}} - \frac{1}{R_{\rm out}})\]

Adopting values typical of the fits in previous sections, 
i.e. $R_{in} = 352 R_s$ 
(corresponding to $T_{\rm in} = 23.6$ eV) 
and $T_{visc} = 7370$ K, we obtain $L_{disc} =  2.2 \times 10^{36} \ergs$.

In the three power laws corresponding respectively to the radio,
NIR and X-ray domains, we integrate $F_{\nu} \propto \nu^\alpha$ 
in the frequency range given in Table \ref{contrib}, 
and multiply it by $4 \pi D^2$ (i.e. assuming isotropic emission).
The corresponding luminosities are given in Table \ref{contrib}. 
Therefore, assuming that $L_{\rm X-rays}$ is representative of 
the contribution from the corona, 
we have $L_{\rm disc} \gtrsim L_{\rm corona}$, and we will discuss this
result later in \ref{juju}.
This is consistent with the fact that 
in the case where all the gravitational power is dissipated in a
static corona, 
nearly half of the coronal luminosity intercepts the disc and is
reprocessed/reflected so $L_{\rm disc} \sim L_{\rm corona}/2$. 
Similarly if a fraction of the
gravitational power is released in the disc (instead of the corona)
$L_{\rm disc} \geq L_{\rm corona}/2$ \citep{haardt:1993}.

\begin{table}
\begin{flushleft}
\begin{tabular}{|c|c|c|c|} \hline
{ }        & { Radio}             & {NIR}                & { X-rays} \\ \hline

%
Frequency  & [9--11.6]            & [11.6--14.9]         &   [14.9--19.3] \\
Luminosity & $4.6 \times 10^{31}$ & $5.3 \times 10^{34}$ & $1.5 \times 10^{36}$ \\
%
\hline
\end{tabular}
\end{flushleft}
\caption[]{\label{contrib} { Contributions $L_{\rm Bol}$ 
in different parts of the SED (called radio, NIR and X-rays following
the convention given in Section \ref{accretion}).
The frequency is given in $\log_{10} (\nu/Hz)$ and the luminosity in 
$\ergs$.} \\
}
\end{table}

        \subsection{The other epochs: evolution of the SED} 
\label{fit_different}

\begin{figure}
\centerline{\psfig{file=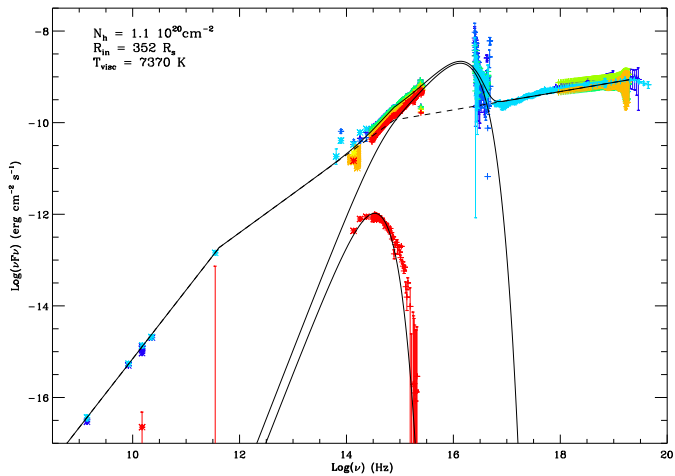,angle=90.,width=9.cm}}
\caption[]{\label{sed_tout+mod} { Spectral Energy Distribution:}
The fluxes are corrected with $N_H = 1.1 \times
10^{20} \cmmoinsdeux$. The overall fit is a multi-colour black
body disc model with an outer disc temperature of 7370 K and inner disc
radius of $352 R_s$.
Straight lines: different power laws, with spectral indices of $0.5$,
$-0.15$ and $-0.8$ (similar to \citealt{hynes:2000}).
The lower curve corresponds to the black-body emission from the
companion star.
}
\end{figure}

\begin{figure}
\centerline{\psfig{file=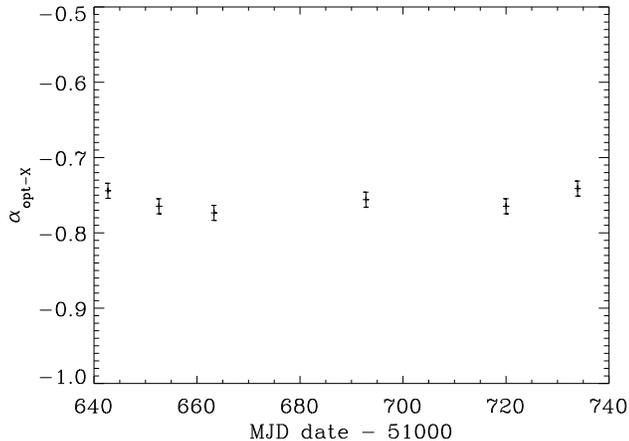,angle=90.,width=9.cm}}
\caption[]{\label{alphaox} Evolution during the outburst 
of the spectral index linking the optical ($10^{14.6}$ Hz) and X-ray 
($10^{18}$ Hz) domains.
}
\end{figure}

There were no gross changes in the SED between epochs 1 and 6, so we 
overplot the fit to epoch 2 with the data from all epochs
in Figure \ref{sed_tout+mod}.
The SED did evolve a little during the outburst, and 
the best way to characterize this evolution is to
quantify the change of the spectral index 
(defined as in Equation \ref{spectral})
between optical ($10^{14.6}$ Hz) and X-ray ($10^{18}$ Hz) domains.
These domains are chosen because both show power-law spectra
and we have simultaneous coverage during the six epochs of observations.
This, reproduced in Figure \ref{alphaox}, shows that the electron
energy distribution remains the same during the whole outburst.
From Figures \ref{sed_tout_radio}, \ref{sed_tout_nir-uv}
and \ref{sed_tout_euv-x} it appears that the fluxes decrease slightly,
however the slopes do not change much.
This suggests that the energy injected in the outflow
decreases during the outburst.

We tried to reproduce this evolution of the SED by modifying 
some parameters of our simple model. One way to do this is by decreasing
the outer (viscous) temperature. However, changing this temperature
modifies the slope of the UV part of the SED, and as we can see
in Figure \ref{sed_tout_nir-uv}, this slope remains the same during
the entire outburst. The other way to act on the multicolour black body
disk would be to modify the inner radius, but this will only 
modify the EUV part of the SED.
Therefore, the only way to reproduce this SED evolution
is to act on the power-laws, and we can indeed do this
just by changing the constant of the power law (and not changing
its exponent). 
This is consistent with a non-thermal contribution which would
have a decreasing energy during the outburst, but the same 
particle energy distribution. It is also suggestive of an outflow
constrained to the central part of the accretion disc, whose size
would decrease during the outburst, which could be correlated with
the increase in QPO frequency during the outburst \citep{wood:2000}.

This evolution is very interesting in the sense that it is different
from what we see in other SXTs, and particularly in other states. 
For instance, in $\xtejdhcn$, \cite{hynes:2002a}
observed a change in the NIR-optical SED that they modelled as
a change in viscous and irradiation temperature (as described in
section \ref{accretion}). 
In contrast, the evolution in 
$\xtejodh$ is mostly due to a change in the non-thermal (outflow) emission,
compared to the thermal (disc) emission.

The lack of pronounced evolution of the SED during 3 months 
is also reminiscent of 
the behaviour of jet sources such as $\grs$ and $\gx$, 
in the so-called ``plateau state'', where only small changes occur
in the lightcurves. This, again, supports the idea of a steady outflow
emanating from the source $\xtejodh$, as we will discuss in
Section \ref{common}.

\section{Discussion} \label{discussion}

        \subsection{Variability and non-thermal contribution} 

A comprehensive discussion about the short time-scale 
variability of $\xtejodh$ 
during all the epochs of observations in the different wavebands 
(including NIR, UV and X-rays) can be found in \cite{hynes:2003}.
Here we just summarize the facts most relevant to our current
analysis by combining our results on the SED and their results on
the variability.
\cite{hynes:2003} show that the XTE data exhibit a 
Poisson subtracted fractional RMS 
of about 35--37\% for the first observations 
decreasing to 26--28\% for the later ones.
The UV variability is weaker than the optical or NIR, and stronger at 
longer wavelengths, with typical rms variability 
$\sim3-4$\,percent in the far-UV and $\sim4-5$\,percent in the 
near-UV. This suggests that at least some of the UV variability 
may be associated with the non-thermal component.

The $\sim 10$ s QPO was seen in the optical, UV and X-rays, implying a common
origin for this QPO throughout the spectrum.
The sampling time of \cite{hynes:2003}'s NIR observations was 
insufficient to search
for the QPO, however they did detect flickering at NIR wavelengths, of larger
amplitude ($\sim 0.8$ mag) than in the optical ($\sim 0.4$ mag).
This NIR variability, up to 50 \%, is
consistent with a dominant non-thermal emission \citep{hynes:2003}. 
Furthermore, the featureless NIR spectrum we took in 2000 June 
(see data plotted in yellow in Figure \ref{sed_tout_decale}) 
is also consistent with the fact
that the disc is not the only source of emission in this part of the spectrum
(for more details concerning the NIR 
see \citealt{chaty:2001d} and \citealt{chaty:2003}).
%
%
Hence both the SED and temporal behaviour 
suggest a strong non-thermal (likely synchrotron) emission at
 radio--UV wavelengths.
Finally, it also shows that the cut-off frequency 
characteristic of the synchrotron
radiation might be in the optical domain or even at higher frequencies.

              \subsection{How common is $\xtejodh$?} \label{common}

                   \subsubsection{SED: Comparison with other galactic sources}

Several aspects of the broad band spectrum of 
this source appear to be very similar 
to those of the other well studied black hole binaries in the low state.
The hard X-ray spectrum with a spectral index 
$\alpha \sim -0.8$ and a cut-off around 100 keV 
is typical of low/hard state spectra observed 
in $\cygxu$ (e.g. Frontera et al. 2001b) 
or $\gx$ (e.g. \citealt{zdziarski:1998}).
$\xtejodh$'s X-ray PDS spectrum is very similar to those of 
other black hole sources \citep{revnivtsev:2000b}. However,
 the characteristic features are significantly shifted towards
 lower frequencies relative to other black hole sources 
with similar masses (cf Figure 2 of \citealt{revnivtsev:2000b}
and Figure 1 of \citealt{sunyaev:2000}).
This suggests that the X-ray emitting region is larger in
$\xtejodh$. 
Therefore, 
 this hard X-ray variability, which is slower than in other SXTs, 
seems in turn to be consistent
 with a particularly large disc inner radius in this source. 
However, such ``slow and ubiquitous'' QPOs have already
been seen in other black hole candidates in the low state. 
For instance, QPOs at $P \sim 20$ s have
been observed both in X-ray and optical in the source $\gx$
\citep{motch:1983}.

Another significant difference with other hard state sources is 
the low inner disc temperature inferred from our data.
In the case of $\cygxu$ in the hard state
 the inner temperature is $kT_{in} \sim 150$ eV 
(see e.g. \citealt{balucinska-church:1995}; 
Frontera et al. 2001b) 
 contrasting with our derived value $kT_{in} \sim 25$ eV (Table \ref{table_nh})
for $\xtejodh$.
This is consistent with a ``very low'' low/hard state \citep{hynes:2000}.
According to the multicolour black body disc model, this lower temperature
implies a larger disc truncation radius than in typical sources
(a few hundreds $R_s$ instead of a few tens).
This can also simply be expressed in terms of $r_{in}/r_{out}$:
this ratio is 4--$7 \times 10^{-5}$ for a typical
SXT like $\xtejdhcn$, which we also observed with similar intensive coverage 
\citep{hynes:2002a}. 
The $r_{in}/r_{out}$ ratio lies 
between $7 \times 10^{-3}$ and $2 \times 10^{-2}$ 
in the case of $\xtejodh$
(depending mainly on the value of the inner radius as derived with our fits).
We also point out that the better energy coverage of the data available 
for $\xtejodh$ helped us getting accurate parameters, since 
the inner disc temperature of hard state sources
is usually poorly determined due to strong interstellar absorption.

Finally, $\xtejodh$ differs from other transient sources 
by its exceptionally low X-ray to optical 
flux ratio \citep{tanaka:1996} as well as its plateau-like light curve 
which contrasts with the exponential decay observed in many transients.
However, once again in order to be consistent we have to compare $\xtejodh$
with other sources in the low/hard state, and when this is done
it becomes less peculiar. For instance, the
X-ray to optical flux ratio in the case of $\gx$ can be as low as
 2.5--3 \citep{motch:1983}, 
very similar to the value of 5 in the case of $\xtejodh$.
Also, its plateau-like lightcurve is common in the jet source
$\grs$ (see e.g. \citealt{tanaka:1996}).

Therefore, the main difference with other sources is
the prominence of the non-thermal contribution.
Indeed, the radio--NIR spectrum presents an inverted spectrum, typical of
non-thermal optically thick synchrotron emission. 
This combined with a lack of pronounced evolution of the SED
strongly represents the signature of a jet.
As a counterpoint, the radio spectral index 
during the 1999 outburst of $\xtejdhcn$ was negative most of the time,
and the interval when it was positive was less than 4 days;
its maximum value was $0.143 \pm 0.180$
\citep{brocksopp:2002}, c.f. the steady $0.5$ value in the
2000 outburst of $\xtejodh$.
The radio-NIR component is much stronger in $\xtejodh$,
exhibiting a low/hard state: in contrast the radio-NIR component
was relatively weak in $\xtejdhcn$'s high/soft state outburst. 
This 
is consistent with the presence of an outflow
(see e.g. \citealt{fender:2001a}).

This signature of a jet 
has been observed many times in $\grs$ (see recent multiwavelength
observations by 
\citealt{ueda:2002}),
$\cygxu$ (see \citealt{stirling:2001} and \citealt{fender:2000a}) 
or $\gx$ \citep{corbel:2000}.
It is usually  
correlated with the appearance of the low-hard X-ray state spectrum,
which is believed to be the result of a coupling between the
Comptonising corona and a compact jet 
(see e.g. \citealt{fender:2001a} and \citealt{corbel:2000}).
We will discuss this in more detail in the Sections \ref{extra} and \ref{jet}.

                        \subsubsection{Comparison with galactic and 
                          extragalactic jet sources} \label{extra}

The strong non-thermal contribution attributed to an outflow, 
implies that $\xtejodh$ is a 
{\it microquasar}.
We therefore compare it to $\grs$, the archetype of the microquasars.
We can also compare $\xtejodh$'s SED
with typical SEDs of {\it quasars} given in \citet{elvis:1994},
following the analysis of \citet{ueda:2002} 
in comparing $\grs$ with typical quasars.

To compare the ratio of energy of the outflow with the accretion energy,
 we need estimates of both.
The outflow energy can be estimated by taking the value 
$\nu F_\nu$ in the radio band ($\nu = 10^{10}$ Hz) for $\xtejodh$,
$\grs$ and quasars.
However, for the measure of the accretion power, we have to
take the wavelength where an optically-thick thermal emission
from the accretion disk presents a peak, therefore 
in the X-rays ($\nu = 10^{18}$ Hz) in the case of $\grs$,
in the EUV ($\nu = 10^{16}$ Hz) in the case of $\xtejodh$
and in the UV ($\nu = 10^{15.2}$ Hz) in the case of quasars.
Then we compare thereafter the ratio of $\nu F_\nu$ 
between the radio and X-rays for $\xtejodh$ and $\grs$ 
with the ratio between radio and UV for quasars.

This outflow/accretion ratio is typically $10^{-6}$ for $\xtejodh$, when it is
between $10^{-7}$ and $10^{-5}$ for $\grs$ respectively
when this source is in a plateau state or exhibiting large radio flares.
The value in $\xtejodh$ is intermediate 
between the two different values exhibited by $\grs$.
The main difference is that $\xtejodh$'s ratio is stable,
instead $\grs$'s ratio is varying on shorter timescales
between the two extreme values.
This suggests that even if the contribution of the outflow to the total
energy budget in the two sources 
is comparable, there is a large difference in their behaviour, 
which is probably due to the fact that $\xtejodh$ is exhibiting
a continuous outflow, and instead $\grs$ shows energetic but
sporadic ejections on short timescales. 
Calculations of energy budget taking into
account different ejection behaviours are given in \citet{chaty:2001a}
by comparing $\grs$ and $\ss$.

For the quasars the outflow/accretion
ratio is typically $10^{-2}$ and $10^{-6}$ respectively
for radio-loud and radio-quiet quasars \citep{ueda:2002}.
This suggests that $\xtejodh$ (and the microquasars) fall into the regime
of radio-quiet quasars. 
The inverted radio spectrum resembles that of radio-quiet quasars, 
and the analogue of the mm-break, around $10^{13}$ Hz, 
exhibited by the radio-quiet quasars and related to self-absorption,
 seems to be the sub-mm excess, around $10^{12}$ Hz, in $\xtejodh$.
On the other hand, if the emission from the outflow also contribute
substantially to the high-energy domain, then the outflow/accretion
ratio becomes even bigger, up to the typical values of radio-loud 
quasars.
If this analogy is real, then the intrinsic luminosity
of the jet could dominate the energetics of $\xtejodh$, as
in the case of the quasars (see Table\,\ref{contrib}).

Therefore, although the high energy spectrum of $\xtejodh$ 
is very similar to what is observed in Seyfert galaxies,
its phenomenology differs widely from that of 
strong extragalactic jet sources such as blazars. 
This is because blazars are dominated 
by relativistic beaming phenomena while the inclination 
in the case of $\xtejodh$ precludes this.

              \subsection{Accretion models}

The low X-ray luminosity as well as the relatively 
large inner disc radius inferred from the comparisons with
the  multicolor blackbody disc model suggests that
either the inner part of the accretion flow (i.e. at radii below $R_{\rm in}$)
is radiatively inefficient 
(efficiency $L_{\rm bol}/(\dot M \times c^2) \sim 5 \times 10^{-4}$), 
or that a large fraction of material is ejected in an outflow, or both.

The ADAF model advocated by \citet{ichimaru:1977} and \citet{narayan:1994}
explains low accretion efficiency as a consequence of
the inner part of the flow advected into the black hole.
The ADAF model was applied to $\xtejodh$ by \cite{esin:2001} who
 found a good fit to the optical--hard X-ray spectrum.
Another possibility could be that the accretion power is advected into a jet 
or outflow, i.e. that
 a significant fraction of the energy could be used to power the jet.
Indeed ADAF solutions imply the flow is not bound 
(c.f. \citealt{blandford:1999}).
This outflow is indeed inferred from the flat radio spectrum
 and analogy with other sources
 in the hard state, where evidence for jets exists,
as we discussed in Section \ref{common}. 
 The most striking example of such sources is
 Cygnus X-1 that presents radio and X-ray spectra very similar to that
 of $\xtejodh$ and  where a milli-arcsec radio jet 
could be resolved \citep{stirling:2001}.
However, since in $\xtejodh$ the fraction of the luminosity that can be firmly 
attributed to the jet represents only a few percent of the total luminosity,
(see Section \ref{contributions}),
 the jet would have to be radiatively inefficient.

Actually the standard 1D ADAF solution
is likely to be strongly affected 
by the presence of the outflow and, more generally
by the development of 
convective instabilities that only appear in 2D solutions
(see e.g. 
\citealt{abramowicz:2000}, \citealt{narayan:2000}, \citealt{stone:1999},
\citealt{igumenshchev:2000}, \citealt{quataert:2000}).
As a consequence of the large theoretical uncertainties, we 
prefer to illustrate the possible 
 presence of a hot optically thin geometrically thick component 
at the center of the cold outer disc using a more phenomenological approach 
based on the constraints given by the observations and energy 
balance conditions as described below.

\subsubsection{The hot disc model} \label{juju}

We will now model the optical--hard X-ray spectrum, by
assuming that the hot component comprises a sphere with radius equal to the
cold disc truncation radius.
The geometry is  very similar to that of the sphere+disc geometry
 detailed in \citet{dove:1997}.

To limit the number of parameters,
 and to avoid the uncertainties about the strength of the 
magnetic field, we further assume that the dominant cooling mechanism 
in the hot plasma 
is Compton cooling by the soft photons produced in the external disc.
Other possible cooling mechanisms
(bremsstrahlung, thermal synchrotron) are assumed to be negligible.
This is a good approximation for luminous
 accreting black hole sources \citep{wardzinski:2000}.
In the case of $\xtejodh$ however, cyclotron/synchrotron and bremsstrahlung 
could be important cooling mechanisms 
(c.f. Frontera et al. 2001a), 
but are not necessarily.
We will discuss this further below, and will show how our assumption
turns out to be consistent with the data.

We therefore used the non-linear Monte-Carlo code of \citet{malzac:2000}
to compute the Comptonised spectrum emitted by the hot phase
(the inner accretion flow) radiatively coupled
 in energy balance with the outer cold standard disc.
In practice, we assume a homogeneous density, dissipation and heating rate 
inside the hot sphere.
For a fixed Thomson optical depth,
 the hot plasma temperature is computed
by balancing  the heating and cooling.
Possible temperature gradients due to the inavoidable inhomogeneity 
of the Compton cooling are accounted for
by dividing the hot sphere in 10
 homogeneous zones with equal volumes
 where the energy balance is computed locally.
We neglect the 
effects of irradiation on the disc temperature profile
 i.e. we assume the standard viscously heated
multi-colour black body disc ($T \propto R^{-3/4}$).
The reprocessed emission is accounted for
by re-emitting the absorbed energy, at the point it impinged
on the external disc, with a black body spectrum at the local disc
temperature. 

The escaping X-ray spectrum is controlled mainly
by the ratio of the volume averaged dissipation
 rate in the hot phase (electron heating) to the
soft flux from the disc that enters the cold phase (controlling the cooling).
This ratio defining the energy balance of the hot component depends both
 on the assumed geometry and  the fraction of radiated power 
dissipated in the hot phase:
\begin{equation}
f=L_{c}/(L_{v}+L_{c})
\end{equation}
where $L_{c}$ is the power radiated in the hot phase, $L_{c} \sim L_{X}$.
$L_{v}$ is the power viscously dissipated in the outer disc.
The cold disc emission arises from both internal viscous dissipation $L_{v}$ but also 
from reprocessing of the Comptonised hard X-rays irradiating the
external disc. For our assumed geometry
 approximately $1/3$ of the hard X-ray luminosity
 is intercepted by the disc. The observed disc luminosity is thus~: 
\begin{equation}
L_{disc}=[(1-f)+f/3] (L_{v}+L_{c})
\end{equation}
and in the case of $\xtejodh$, $f$ can be determined observationally:
\begin{equation}
f\sim1/(\frac{2}{3}+ \frac{L_{disc}}{L_{X}})\sim 0.33
\end{equation}

The fraction of disc luminosity
due to reprocessing is small, $\sim  \frac{f}{3-2f} = 0.14$ and irradiation 
is unlikely to affect the outer disc structure.
In particular, the local temperature depends only weakly
 on the total flux ($T \propto F^{1/4}$), thus  
 for a local irradiating flux of 20 \%
of that of the local viscous flux,
the disc temperature increases by less than 5 \%.
We note that this is perfectly in agreement with our observations
suggesting that effects of disc irradiation on the temperature profile
 should be weak (if any, see Section \ref{irradiation}). 
In the framework of the sphere+disc model, 
this ``weak irradiation'' hypothesis is corroborated by the relatively
low observed ratio of Comptonised to thermal emission.
Therefore our simple multi-colour black body and power
law fits are consistent with our more sophisticated model assumptions.
We also note that this geometry predicts an amplitude for the reflection 
component 
$R\sim 0.3$ which is slightly above the upper limits obtained
 by Frontera et al. (2001a) 
for their fits assuming low metallicity 
abundances in the disc (but see also \citealt{miller:2002}). 
However, even if
such low Z abundances might be related to a possible 
halo origin, as argued for by \cite{wagner:2001} and \cite{mirabel:2001}, 
we point out that the metallicity
should be higher than expected for a halo object,
since the mass donor star is probably a CNO processed core 
\citep{haswell:2002}.

The spectral shape depends on 5 parameters:
 the disc outer temperature $T_{out}$
and radius $R_{out}$, the disc inner radius $R_{in}$,
 and the fraction of accretion energy
dissipated in the hot sphere to that dissipated in the accretion disc $f$, 
the hot sphere optical Thomson depth $\tau$ with respect to the sphere radius.
The unique broad-band coverage in $\xtejodh$ constrains all
 the parameters relatively well.

\begin{figure}
\centerline{\psfig{file=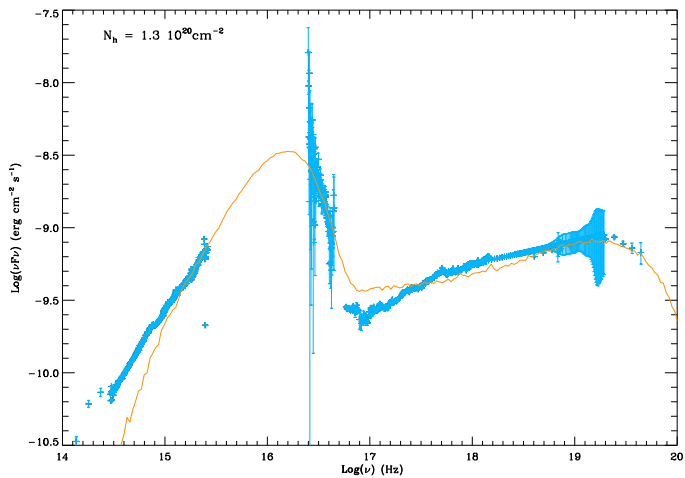,angle=90.,width=9.cm}}
\caption[]{\label{sed_vis2_juju} Modelling the optical to 
hard X-ray spectrum. 
The hot disc scheme is illustrated here, 
where we compare the non-linear Monte-Carlo simulation
 for a central spherical 
hot plasma which radiates through Comptonisation of the soft photons 
emitted by 
the outer cold disc both through internal viscous dissipation and reprocessing 
of the Comptonised hard X-rays
(the later being $\approx 20$ \% of the disc emission).
See section \ref{juju}.
}
\end{figure}

In the simulation shown in Figure \ref{sed_vis2_juju}
the cold disc parameters 
(and especially the locally emitted black body spectrum at its surface
corresponding to the standard temperature dependence) 
have been set to values close
to that assumed or derived from the fits of section \ref{fit_epoch2},
namely $R_{out}= 12 \times 10^3$ $R_{s}$ (80 \% 
of the Roche lobe radius), $R_{in}=$300 $R_{s}$, $T_{out}=8000$ K.
The Thomson optical depth along the sphere radius was fixed at
$\tau=1$ as indicated by the fit of 
Frontera et al. (2001a) 
for a spherical geometry, and $f=0.33$ as discussed above.
We did not include in the model any absorption as seen
at $\log(\nu) \sim 17$ which was attributed to
a warm absorber by \cite{esin:2001}. 

The most important parameters for the X-ray emission are $f$ 
and $\tau$. $f$ controls the Compton $y$ parameter of the plasma 
at equilibrium $y \propto \tau T_{e}$
 and thus the X-ray photon index $\Gamma$ is very sensitive 
to $f$.
 With the $f$ value independently
 provided by the observed luminosity of the cold and hot 
components, the model produces $\Gamma=1.82$ (2.-10 keV),
 in agreement with the results
 from spectral fits of the {\it SAX} data 
(Frontera et al. 2001a). 
Since all the parameters determining the X-ray slope are independently
 constrained, such an agreement is very remarkable.
At fixed $f$, the parameter $\tau$ controls mainly
 the temperature of the hot phase. From energy balance we
 get a volume averaged equilibrium temperature of the sphere of $kT_{e}=
$ 109 keV in agreement with the value inferred from the fits of 
{\it SAX} data (Frontera et al. 2001a). 
As a consequence, the shape of the high energy cut-off is well reproduced. 

Note that this model differs from the ADAF model of \cite{esin:2001}
 who considered cyclo-synchrotron radiation 
as the main source of soft photons for Comptonisation and neglected 
the soft-photons from the disc. 
We did precisely the opposite.
In our model, if the bremsstrahlung and cyclo-synchrotron radiation 
were to constitute a significant additional source of soft photons, this
 would increase the cooling rate in the hot plasma, and,  as a result 
we would get a much softer spectrum than was observed.
The observations would then require a different geometry, reducing 
the soft photon input from the outer disc. 
For example an oblate geometry for the central hot region could satisfy
 this requirement.
In the extreme limit where the hot phase is geometrically thin along the disc
 axis we would be in the situation 
considered by \cite{esin:2001}, dominated by cyclo-synchrotron radiation.
Note that the correlated variability between the optical and X-rays suggests 
a causal connection, and 
that the synchrotron radiation emanating from the same region 
that produces the X-rays can contribute in the optical band 
\citep{merloni:2000}, but 
not necessarily as a significant source of soft cooling photons.
Therefore this processus is not inconsistent with our model where we
neglected the synchrotron emission.

\subsubsection{jet corona models} \label{jet}

If the X-ray emission is produced in the
central part of the accretion flow, the $\sim 0.1$ Hz QPO
 observed both in the X-ray and optical 
 is difficult to explain \citep{merloni:2000}.
On the other hand it may be explained simply by
 a flux modulation at the Kepler frequency
 of the disc at $R_{in}$.
This suggests that a non-negligible fraction of the flux would
 be produced in a transition region where both the hot plasma and cold 
thin disc coexist.

This transition region could consist of a hot corona  atop the external disc.
 The generation of magnetic
 field in the disc through magnetorotational instability and the subsequent 
buoyancy of the magnetic field 
(\citealt{tout:1992}, \citealt{miller:2000})
 is generally invoked as an efficient 
mechanism for the transfer of magnetic energy from the disc to a corona 
where it is dissipated through magnetic flares. 
This scenario for energy dissipation in the corona was built to explain
 a strong coronal emission in the innermost part of the accretion flow.
At the large distances we infer for the disc truncation radius, 
it is expected to be less efficient.

Therefore, if $R_{in}$ is as large as we infer,  
 alternative dissipation processes such as transport and heating through 
Alfv\'en waves \citep{tagger:1999} or viscous dissipation
 in a fast accreting corona \citep{rozanska:2000} may appear preferable. 
Actually, if the inner flow was very inefficient (or even nonexistent), 
the X-ray emission could be fully dominated by the coronal emission.
In this context, 
 mildly relativistic ejections of coronal plasma are likely to be important 
in hard state sources (\citealt{beloborodov:1999}, \citealt{malzac:2001a}). 
In particular it may 
explain the relatively hard X-ray spectrum and absence of reflection
observed in $\xtejodh$ together with being consistent with the presence 
of an outflow. 
From the theoretical side relativistic outflow models often require
 an accretion disc corona as the place where the flow is initially powered 
(see \citealt{tagger:1999}, \citealt{merloni:2001}).
The accretion disc corona picture would be thus physically compatible
 with the presence of a jet inferred from the flat radio spectrum.

In accretion disc corona models a substantial part of the disc emission
 is due to reprocessing of the hard X-rays impinging on the disc.
 As the optical, near-UV and far-UV variability indicates that reprocessing is
weaker than the extremely strong synchrotron component 
 \citep{hynes:2003}, this constrains the corona to be active
 only in the innermost part of the disc. 
The X-ray luminosity should then come from a region of the disc forming
 a thin ring with inner radius $R_{in}$ and outer radius $R_{c}$. 

One can get an estimate of $R_{c}$ by assuming 
that at distances lower than $R_{c}$, all the accretion power
 goes to the corona while at larger distance it is dissipated viscously
 in the disc.
Since the estimated X-ray luminosity is about half of that of the disc and  
about half of the coronal emission is intercepted by the disc
 we can conclude that about half of the disc luminosity
 could be due to the reprocessing.
Assuming that reprocessed luminosity is half 
of that of the disc one gets $r_c \sim 2 R_{\rm in}$.

We point out that our results are perfectly consistent with
the Accretion-Ejection Instability (AEI) model \citep{tagger:1999}
which predicts that Alfv\'en waves 
efficiently power the corona and the jet in a range of distance
 comprised between $R_{\rm in}$ and the corotation radius of the Rossby 
vortex at a few $R_{\rm in}$ \citep{varniere:2002b}.
Particularly, from the value of the inner radius that we derive 
with our fits, the AEI predicts a QPO which is well in agreement
\citep{varniere:2002}
with the values observed in UV, optical and X-rays 
during the outburst at Fourier frequencies $\sim 0.1$ Hz.
The prediction on the QPO frequency evolution
would also be consistent with the observations by \cite{wood:2000}.
The AEI context, by assuming equipartition between magnetic
field and gas pressure, would exclude a Schwarzschild black hole,
and favor a Kerr black hole rotating at
a spin between 0.90 and 0.99 (P. Varni\`ere, priv. com.).
This QPO  may be explained simply by a flux modulation 
due to the keplerian rotation of the disc or, in the \cite{tagger:1999}
 framework, the presence of the spiral wave (P. Varni\`ere, priv. com.). 

We also note that in addition to the Comptonised emission from the hot
 disc and/or corona, a synchrotron component from the jet could also
contribute in the X-rays. Such a non-thermal component was seemingly
detected in $\grs$ in a low/hard state \citep{vadawale:2001}.
In fact, the whole X-ray emission could even arise from pure synchrotron
from the jet as demonstrated by \citet{markoff:2001}.
This model attributes all the X-ray luminosity of $\xtejodh$ to
synchrotron emission from the jet. A potential  problem
 for this model is that it predicts essentially 
no reflection component.
This makes it difficult to transpose to other similar 
jet sources such as $\cygxu$ 
where a significant reflection component is clearly observed and
correlated with other X-ray spectral characteristics such as the photon
index \citep{gilfanov:1999}.
However the pure synchrotron model cannot be formally ruled out,
and we could also have both mechanisms acting at the same time.

\section{Conclusions}

We reported multiwavelength observations of $\xtejodh$ during its
outburst, assembling the most complete spectral energy distribution (SED) 
of this source yet published, including our observations with UKIRT, {\it HST}, 
{\it RXTE}, {\it EUVE}, and adding observations from the literature:
Ryle Telescope, VLA, JCMT, {\it Chandra} and {\it SAX}.
We followed the source for 6 months, and show its evolution
during the outburst.
The main results of our broadband multi-epoch coverage are:

(1) The source $\xtejodh$ was in a very low state
throughout the outburst (estimated inner radius at $350 R_s$).

(2) The column density is low, 
between $0.80$ and $1.30 \times 10^{20}\cmmoinsdeux$.

(3) The accretion disc seems to be heated mainly by viscosity throughout
 the outburst, without a strong irradiation contribution.

(4) It exhibited an inverted spectrum 
from radio to at least optical wavelengths, 
characteristic of a strong non-thermal (likely synchrotron) contribution,
usually attributed to an outflow.

(5) By examining the near-quiescent system we found that
the best-fit parameters were 
a fractional mass donor star contribution of $25 \pm 2 \%$,
a remnant accretion disc at $6000 \pm 50 K$
and a distance of $1.71 \pm 0.05 \kpc$.

(6) The quasi-absence of variation of the SED during 3 months 
is consistent with a steady outflow emanating from the source,
similarly to other ``microquasars'' in the ``plateau'' state.

(7) The ratio between radio and EUV energy, indicative
of the outflow to accretion energy ratio, suggests that the source falls
into the regime of the radio-quiet quasars, and is also consistent with
a steady outflow.

(8) The small change of the SED in the optical--UV part,
along with the constant power-law slopes during the whole
outburst, is consistent with a continuous, gradual
 decrease of the outflow energy.

(9) We modelled the emission from the optical to the hard X-rays
with a hot disc model, showing that 
the high-energy part of the spectrum can emanate from the
accretion flow.
We therefore have to take into account the possibility
 that the high-energy emission
from this source comes from i) Comptonization ii) pure synchrotron
or iii) a mixture of both.

Although this object exhibits peculiar characteristics,
e.g. a large inner disc radius and a likely origin in the halo,
some of its characteristics are
very similar to other sources, particularly in the low--hard state
and exhibiting jets or outflows.
This object is particularly important
because its very low absorption reveals phenomena
that until now were difficult to study.
$\xtejodh$ facilitates testing and refinement of models 
for variability and emission in black hole accretors.

\section{acknowledgements} 

We thank Christopher W. Mauche for fruitful discussions about
absorption in the EUV, and for giving
us the {\it EUVE} data.
We are grateful to Mike Garcia for useful discussions about the evolution
of the SED, to him and Jeff McClintock for 
giving us the {\it Chandra} data, 
and to Filippo Frontera for the {\it SAX} data.
We thank Tom R. Geballe who obtained the 
NIR spectrum of $\xtejodh$ on 2000, June 26.
We thank Guy Pooley for the Ryle Telescope data also used in this Figure,
and VSNET for all their alerts on $\xtejodh$ and their optical data 
used in Figure \ref{1118_lc}. 
S.C. and J.M. thank Michel Tagger, Peggy Varni\`ere and Josep Mart$\ie$.
%
S.C., C.A.H. and R.I.H. gratefully acknowledge
support from grant F/00--180/A from the Leverhulme Trust.
S.C. and J.M. acknowledge a travel grant from the Groupe de Recherche
{\it Ph\'enom\`enes Cosmiques de Haute \'Energie} of the French Centre
National de la Recherche Scientifique.
J.M. acknowledges a grant from 
the European Commission (contract number ERBFMRX-CT98-0195,
 TMR network ``Accretion onto black holes,
compact stars and protostars'').
W.C. also acknowledges support from HST-GO-08647.10-A.

The United Kingdom Infrared     Telescope is operated by  the    Joint
Astronomy Centre on behalf of the  U.K. Particle Physics and Astronomy
Research Council.  UKIRT  Service  observations of  $\xtejodh$ through
the year    2000 were obtained   thanks  to  override time   which was
pre-approved in case of outbursting transients (U/00A/45, PI S.C.), to
be coordinated   with our  {\it   HST}  and  {\it  RXTE} observations.
S.C. is grateful to the UKIRT staff  
for these override service  observations, and  in
particular to Andy  Adamson, John K.  Davies, Sandy K.  Leggett and Chris
Davis.

Based on observations made with NASA/ESA Hubble Space Telescope, 
associated with proposal GO 8647. 
Support 
was provided by NASA through a grant from 
the Space Telescope Science Institute, which is operated by the Association 
of Universities for Research in Astronomy, Inc., under 
NASA contract NAS 5--26555. 
We finally thank the {\it HST/STScI} and {\it RXTE} support staff 
for ongoing efficient
effort in these multi-epoch campaigns.
This research has made use of NASA's Astrophysics Data System 
Bibliographic Services and quick-look results provided by the ASM/RXTE team.


\end{document}